\documentclass{kapproc} 
 \usepackage{procps}
\usepackage{graphicx}

\let\footnote\savefootnote
\let\footnoterule\savefootnoterule

\def\eq#1{\begin{equation} #1 \end{equation}}
\def\E#1{\hbox{$10^{#1}$}}
\def\sub#1{_{\rm #1}}
\def\about  {\hbox{$\sim$}}
\def\ga     {\mathrel{\hbox{\raise.3ex\hbox{$>$}\llap
                                {\lower.8ex\hbox{$\sim$}}}}}
\def\la     {\mathrel{\hbox{\raise.3ex\hbox{$<$}\llap
                                {\lower.8ex\hbox{$\sim$}}}}}
\def\x      {\hbox{$\times$}}
\def\mic    {\hbox{$\mu$m}}
\def\Lo     {\hbox{$L_\odot$}}

\def\Mdot   {\hbox{$\dot M$}}
\def\Myr    {\hbox{$M_{\odot}$ yr$^{-1}$}}
\def\QV     {\hbox{$Q\sub V$}}
\def\Qast   {\hbox{$Q_\ast$}}
\def\tV     {\hbox{$\tau\sub V$}}
\def\tF     {\hbox{$\tau\sub F$}}
\def\vf     {\hbox{$v_\infty$}}
\def\vfff   {\hbox{$v_\infty^3$}}
\def\kms    {\hbox{km s$^{-1}$}}
\def\ss     {\hbox{$\sigma_{22}$}}
\def\H2O{\hbox{H$_2$O}}

\begin{document}

\articletitle{The Structure of Winds in AGB Stars}

 \author{Moshe Elitzur}
 \affil{Department of Physics \& Astronomy\\
  University of Kentucky, Lexington, KY 40506}
 \email{moshe@uky.edu}

 \author{\v Zeljko Ivezi\'c}
 \affil{Department of Astrophysical Sciences\\
  Princeton University, Princeton, NJ 08544}
 \email{ivezic@astro.princeton.edu}

 \author{Dejan Vinkovi\'c}
 \affil{Department of Physics \& Astronomy\\
  University of Kentucky, Lexington, KY 40506}
 \email{dejan@pa.uky.edu}

\begin{abstract}
Most dusty winds are described by a set of similarity functions of a single
independent variable that can be chosen as \tV, the overall optical depth at
visual. The self-similarity implies general scaling relations among the system
parameters, in agreement with observations. Dust drift through the gas has a
major impact on the structure of most winds.
\end{abstract}


\section{Introduction}

The complete description of a dusty wind should start with a full dynamic
atmosphere model and incorporate the processes that initiate the outflow and
set the value of \Mdot. These processes are yet to be identified with
certainty, the most promising are stellar pulsation (e.g. Bowen 1989) and
radiation pressure on the water molecules (e.g. Elitzur, Brown \& Johnson
1989). Proper description of these processes should be followed by grain
formation and growth, and subsequent wind dynamics. Two ambitious program
attempting to incorporate as many aspects of this formidable task as possible
have been conducted over the past few years by groups at Berlin (see J.M.\
Martin, these proceedings) and Vienna (see Dorfi et al 2001). While much has
been accomplished, the complexity of this undertaking necessitates
simplifications such as a pulsating boundary. In spite of continuous progress,
detailed understanding of atmospheric dynamics and grain formation is still far
from complete.

Fortunately, the full problem splits naturally to two parts, as recognized long
ago by Goldreich \& Scoville (1976). Once radiation pressure on the dust
exceeds all other forces, the rapid acceleration to supersonic velocities
decouples the outflow from the earlier phases---{\em the supersonic phase would
be exactly the same in two different outflows if they have the same mass-loss
rate and grain properties even if the grains were produced by entirely
different processes.} Furthermore, these stages are controlled by processes
that are much less dependent on detailed micro-physics, and are reasonably well
understood. And since most observations probe only the supersonic phase, models
devoted exclusively to this stage should reproduce the observable results while
avoiding the pitfalls and uncertainties of dust formation and the wind
initiation.

\section{Dusty Winds: Problem Setup and Solution}

A steady-state dusty wind is  controlled by three forces. First, radiation
pressure drives the outflow. When $\tV > 1$, the radiation is continually
reddened and its coupling to the dust is degraded; this weakening of the force
can be described by a radial profile $\phi(r)$. Radiation pressure acts mostly
on the dust grains while the wind mass is mostly in the gas particles, and
collisional coupling between these two components introduces the second force,
drag. Gilman (1972) has shown that the dust-gas relative velocity reaches
steady state within a distance much shorter than all other scales. The drag
force can then be eliminated, because the dust fully mediates to the gas the
radiation pressure force. But the drift still has an important effect, since
the radiative force per unit mass is proportional to $n_d/n$, and separate mass
conservation for the dust and gas implies $n_d/n \propto v/v_d$.\footnote{Even
with prompt dust formation and no further grain growth or destruction, the dust
abundance varies in the shell because of the difference between the dust and
gas velocities.} Finally, the gravitational pull of the central star opposes
the expansion and must be overcome in any outflow.

Accounting for the three forces, the equation of motion in terms of
dimensionless velocity $w$ and radial distance $y$ is (Elitzur \& Ivezi\'c
2001)
\eq{\label{eq:basic}
 {dw^2\over dy} =
   {P^2\over y^2}\left(\phi(y)\zeta(y) - {1\over\Gamma}\right).
}
The constants $P$ and $\Gamma$ are the ratios of radiation pressure to the drag
and gravity, respectively, $\zeta = v/v_d$ is the drift profile and $\phi$ is
the reddening profile, obtained from the separate, but coupled, equation of
radiative transfer. The outflow problem is fully described by this simple
mathematical equation with only three free parameters---the constants $P$ and
$\Gamma$ and the initial value $w(y = 1)$. There are no other free parameters.
The wind structure will be exactly the same in systems that have different mass
($M$), luminosity ($L$) or mass loss rate (\Mdot) when these properties combine
to produce the same $P$ and $\Gamma$.

The complete solution of equation \ref{eq:basic} is presented in Elitzur \&
Ivezi\'c (2001), including the analytic solution for optically thin winds ($\tV
< 1$). Here we present observational implications.

\subsection{Physical Domain}

The relation $\Mdot\vf \le L/c$ has often been used as a physical bound on
radiatively driven winds, even though the mistake in this application when $\tV
> 1$ has been pointed out repeatedly (e.g.\ Ivezi\'c \& Elitzur 1995). Instead,
the proper form of momentum conservation is $\Mdot\vf = \tF L/c$, where \tF\ is
the flux-averaged optical depth, and it yields
\eq{\label{eq:momentum}
 \Mdot\vf = {L\over c}\left(\Qast/\QV\right)\,\tV\left(1 + \tV\right)^{-0.36}.
}
Here \QV\ is the 
dust efficiency coefficient at visual and \Qast\ is its Planck-average with the
stellar temperature. Typical values of \Qast/\QV\ are 0.1 for silicate and 0.25
for carbon dust. Since the ratio $\Mdot\vf/(L/c)$ can exceed unity and increase
without bound, momentum conservation does not constrain the wind. Instead, the
constraints come from force considerations---the outward force must exceed
gravity (positive derivative in eq.\ \ref{eq:basic}), leading to the two phase
space boundaries shown in figure 1. The liftoff bound, shown in the left panel,
sets a lower limit on \Mdot, proportional to $M^2/L$; below this minimal \Mdot\
the grains are ejected without dragging the gas because the density is too low
for efficient dust--gas coupling. As a condition on the wind origin, this bound
is the most uncertain part of the solution. The result $\Mdot_{min} \propto
M^2/L$ is reasonably secure (a similar relation was noted by Habing et al
1994), but the proportionality constant involves a fudge factor. This factor
can only be determined from a more complete formulation that handles properly
grain growth. The bound shown in the right panel reflects the weakening of
radiative coupling in optically thick winds because of the radiation reddening,
setting an upper limit on \Mdot.

\begin{figure}
\includegraphics[width=0.5\hsize,clip]{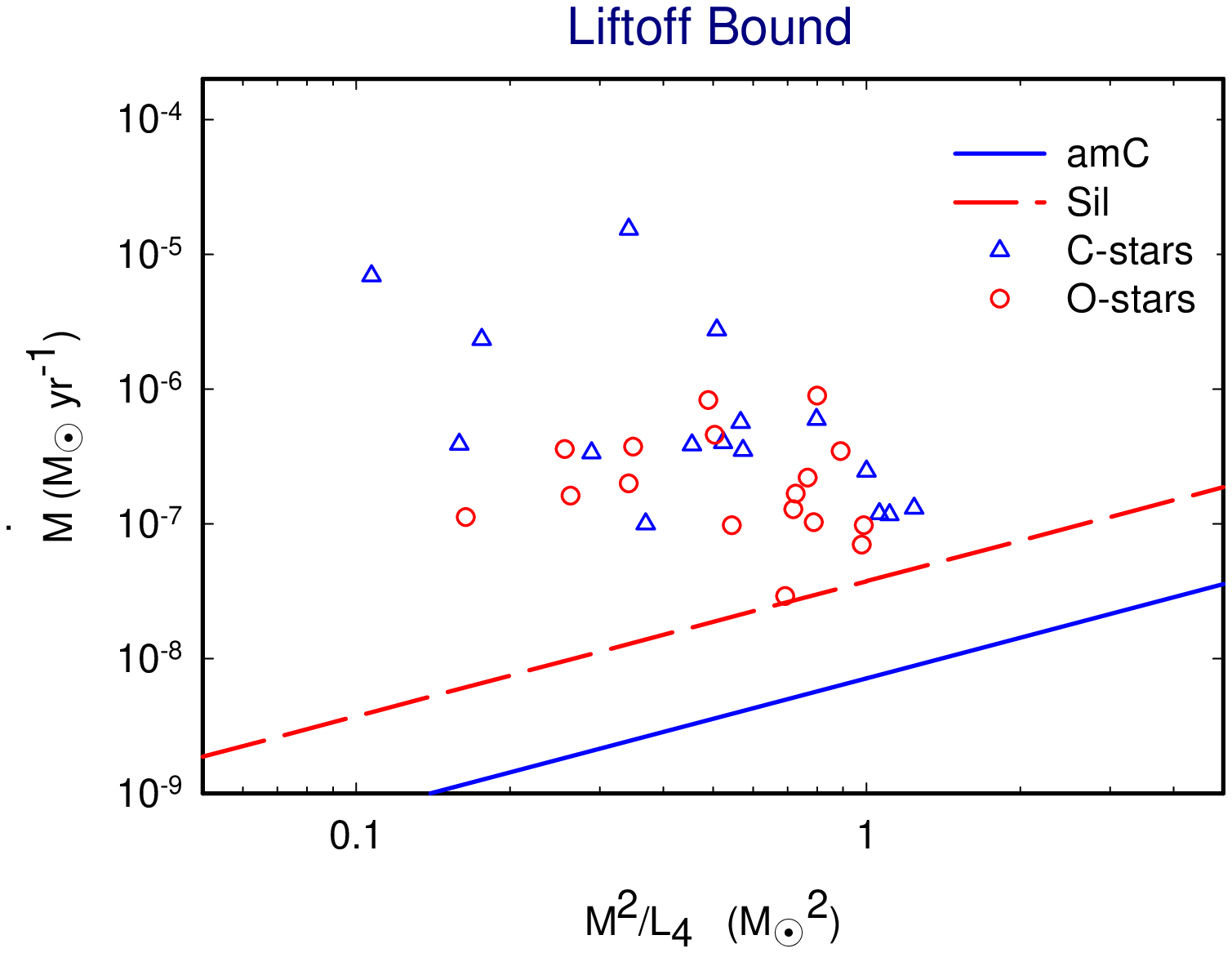} \hfil
\includegraphics[width=0.5\hsize,clip]{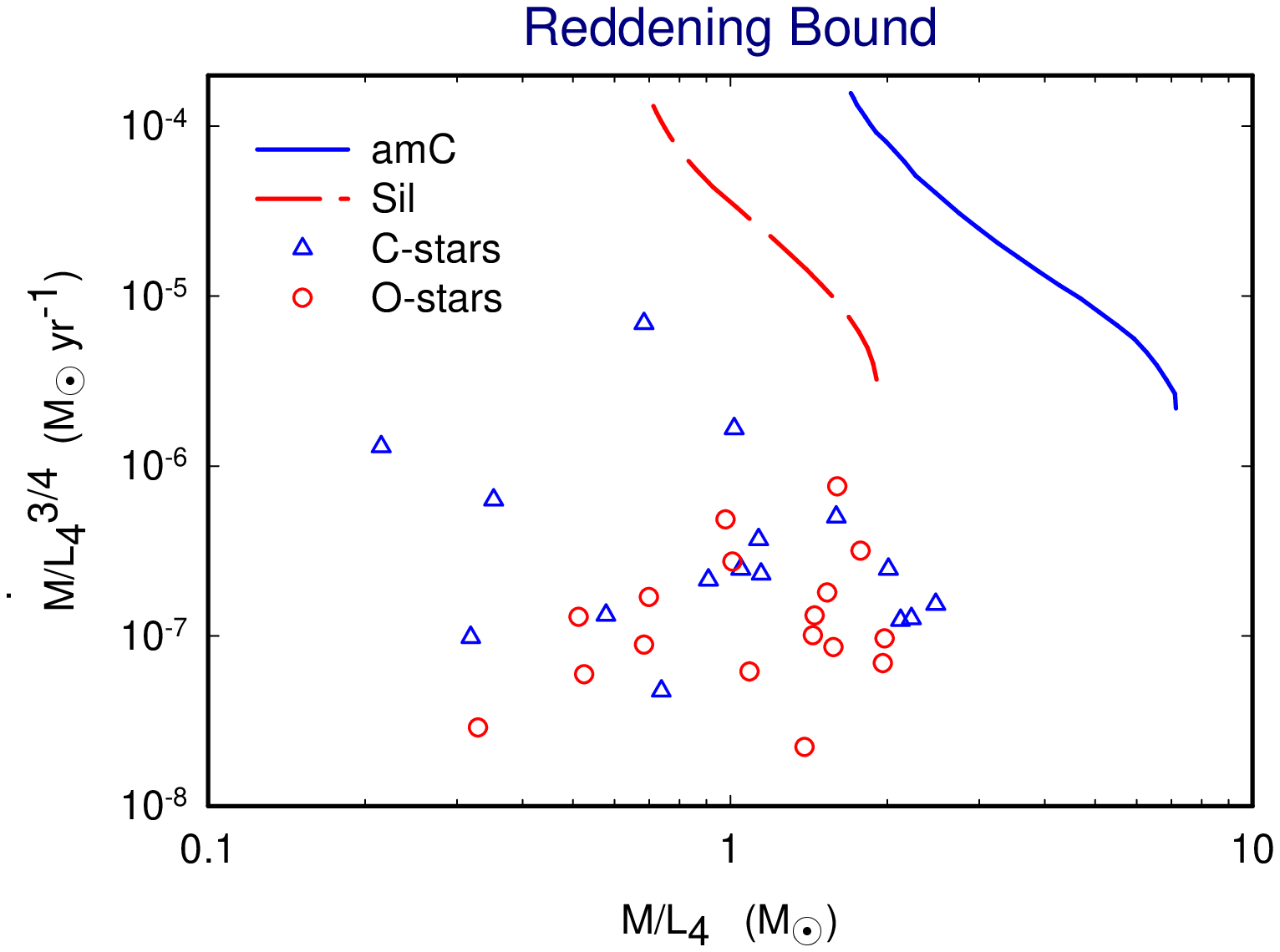}
\caption{Winds powered by radiation pressure on dust should fall above the
liftoff bounds shown in the left panel and below the reddening bounds in the
right one. $L_4$ is the stellar luminosity in \E{4} \Lo\ (Ivezi\'c \& Elitzur
2002).}
\end{figure}

\subsection{Scaling}

Most stars are located well inside the allowed region of phase space, as is
evident from figure 1. The dependence on $\Gamma$ and $w(1)$ disappears rapidly
with distance from the boundaries, therefore most dusty winds are described by
a set of similarity functions of just a single independent variable $P$, which
is equivalent to \tV. The self-similarity implies general scaling relations.
Systems with the same combination $\Mdot/L^{3/4}$
necessarily have also the same $\Mdot\vf/L$ because both are uniquely related
to \tV.

The velocity profile for all winds can be summarized with the simple expression
\eq{
    v = \vf\left(1 - {r_a\over r}\right)^k,     \qquad \hbox{where}\
  \cases{k    =   \frac23  & when $\tV \la 1$ \cr
         k \simeq    0.4   & when $\tV \ga 1$ \cr}
}
and where $r_a = r_1[1 - (v_1/\vf)^{1/k}]$, with $v_1$ the wind velocity at its
starting radius $r_1$.  The profile shape of $v/\vf$ is nearly the same for all
winds, a weak dependence on \tV\ enters only in the power $k$. The small-\tV\
value $k = \frac23$ reflects the effect of the drift, which dominates the
dynamics in that regime, the switch to a more moderate profile at large \tV\ is
caused by reddening.

The dimensionless solution involves only properties of individual grains. The
dust abundance is irrelevant and affects only the velocity scale \vf. When only
the radiation pressure force is taken into account, \vf\ increases with $L$ and
is independent of \Mdot\ (Goldrecih \& Scoville 1976). Drift effects change
this behavior fundamentally, producing instead
\eq{\label{eq:Young}
            \vfff = A\,\Mdot\left(1 + \tV\right)^{-1.5},
}
where the proportionality coefficient $A$ contains properties of the grain
material and $n_d\sigma_d/n$. Therefore, outflows with $\tV < 1$ should obey
$\vfff \propto \Mdot$ if they have the same dust properties. Remarkably, {\em
even though the wind is driven by radiation pressure, its velocity is
independent of luminosity}. The dependence on $L$ enters only in optically
thick winds because $\tV = A^{1/3}(\QV/\Qast)\Mdot^{4/3}\!/L$.

The behavior implied by eq.\ \ref{eq:Young} had been noticed by Young (1995)
from 36 nearby Mira variables with low mass-loss rates. Young's data is
presented in figure 2 as a histogram of \vfff/\Mdot\ together with a similar
histogram for C-rich stars with small optical depths (data from Olofsson et al
1993). Each histogram shows a pronounced peak---as implied by eq.\
\ref{eq:Young} when the dust properties do not vary among members of the set.
The value of $n_d\sigma_d/n = \E{-22}\ss$ cm$^{-2}$ determined for each set
from its average $A$ and properties of silicate and carbon grains is listed in
the figure. Remarkably, the two samples of different types of stars produce the
same \ss. In spite of the large differences in atmospheric and grain properties
between O- and C-rich stars, the fraction of material channelled into dust is
such that \ss\ turns out to be the same in both.

\begin{figure}
\includegraphics[bb=83 421 516 761,width=0.5\hsize,clip]{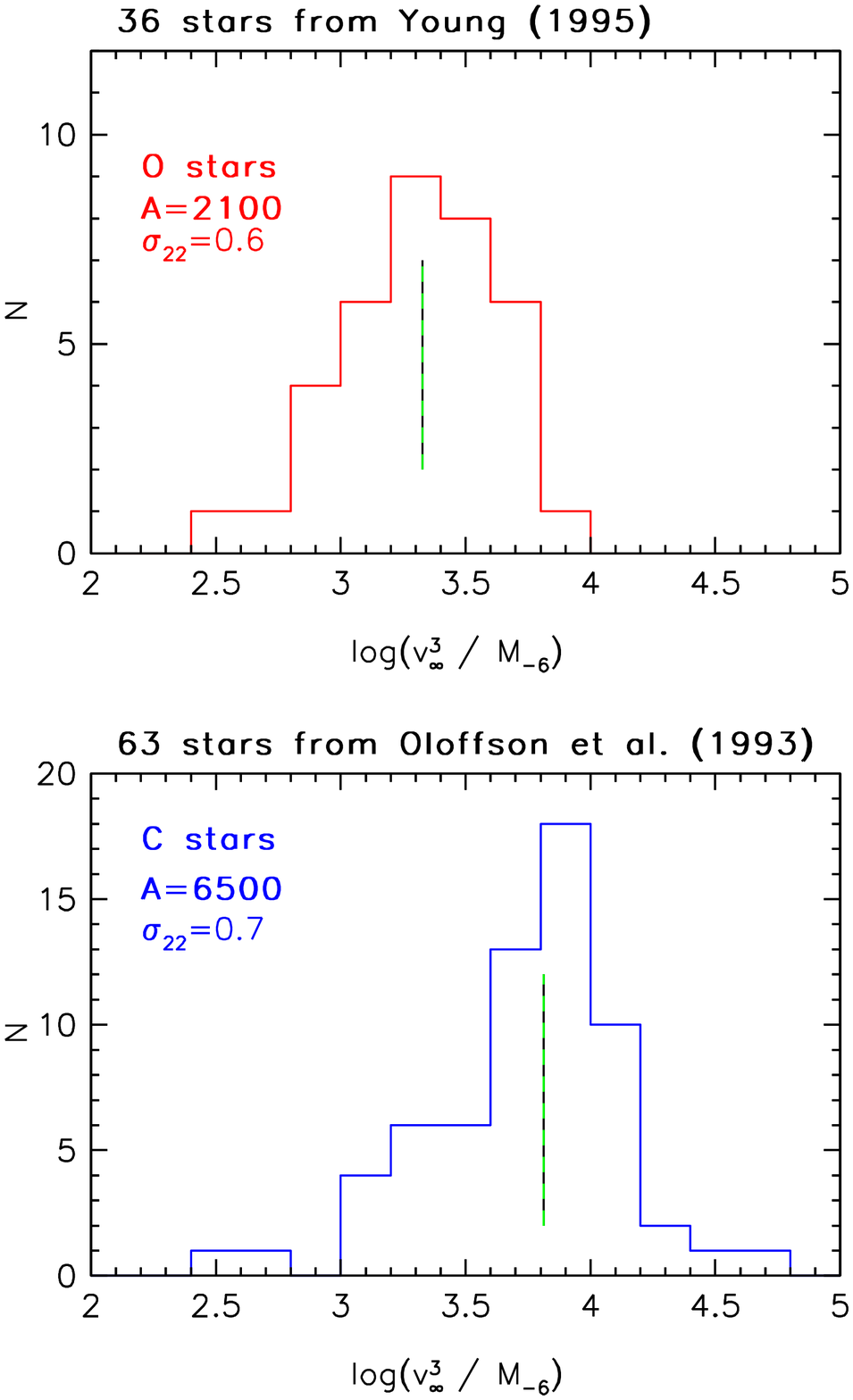} \hfil
\includegraphics[bb=83  60 516 398,width=0.5\hsize,clip]{moshe_fig2.ps}
\caption{Histograms of \vfff/\Mdot\ for oxygen- and carbon-rich stars with $\tV
< 1$ outflows.  In each data set, both \vf\ (in \kms) and \Mdot\ (in \E{-6}
\Myr) were determined from CO observations. A vertical line marks the mean of
each distribution. The corresponding coefficient $A$ (see eq.\ \ref{eq:Young})
is listed together with the value of \ss\ it implies (Ivezi\'c \& Elitzur
2002).}
\end{figure}

Drift and reddening play a major role in the dynamics. As \Mdot\ decreases, the
dust and gas decouple and the velocity decreases too. As \Mdot\ increases and
the wind becomes optically thick, reddening degrades the efficiency of the
radiation pressure force and the velocity again decreases. In between, the
velocity reaches maximum $v\sub{max} \simeq 20L_4^{1/4}$ \kms, obtained at
$\Mdot(v\sub{max}) \simeq 2\x\E{-6}L_4^{3/4}$ \Myr. The proportionality
coefficients are slightly different for carbon and silicate dust.

\begin{figure}
\includegraphics[width=0.50\hsize,clip]{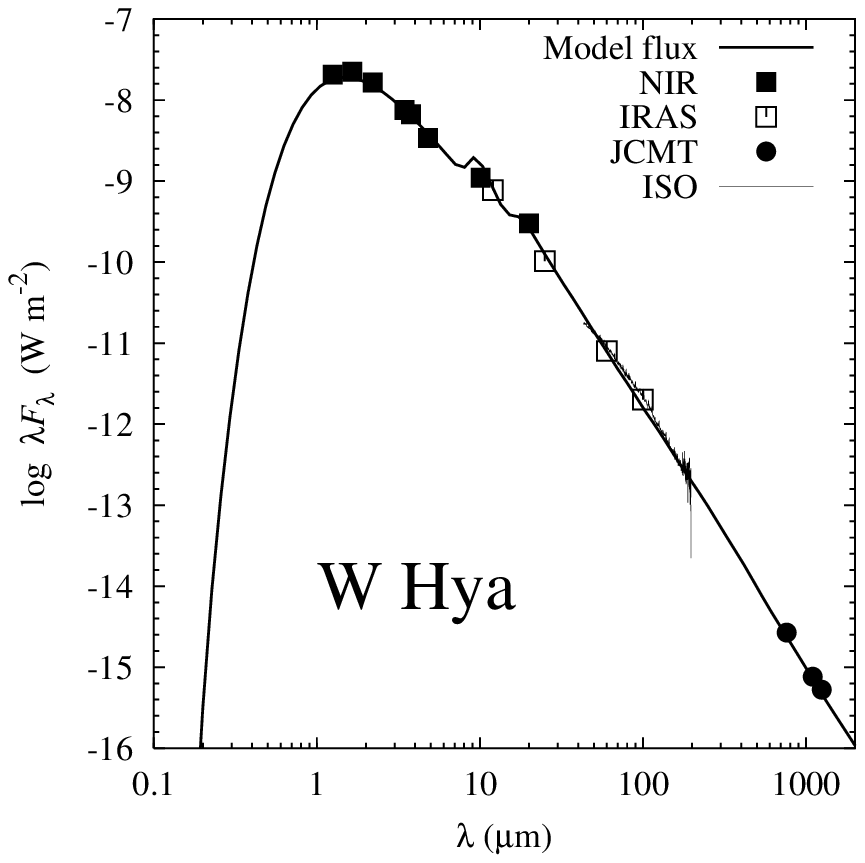} \hfil
\includegraphics[width=0.50\hsize,clip]{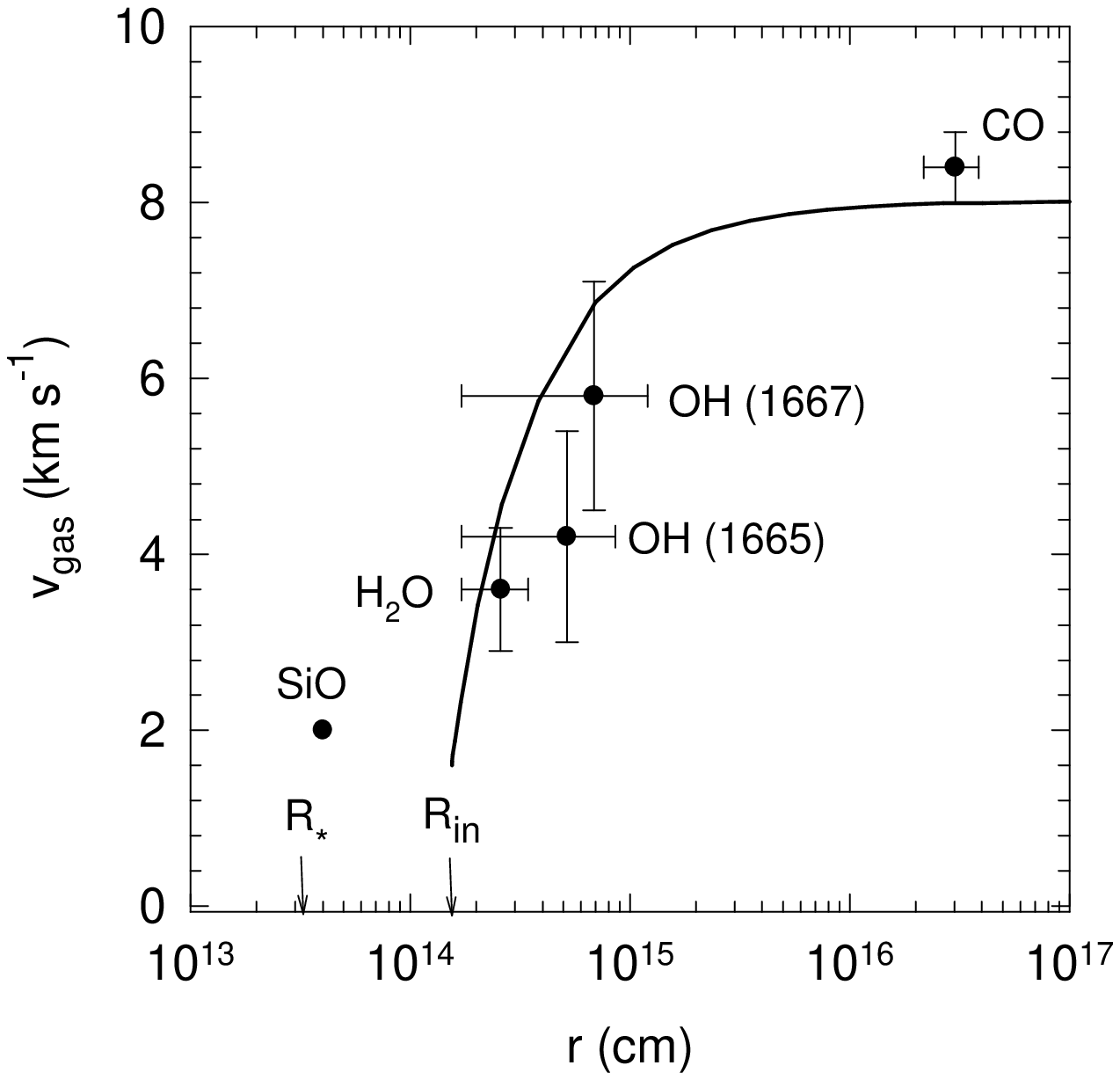}
\caption{Modeling the wind in W Hya (Zubko \& Elitzur 2000) with outflow
calculations by the code DUSTY (Ivezi\'c et al 1999). Only one free parameter
was varied, \tV, determined to be 0.83 from the SED fit in the left panel.
Right: Model prediction for the gas velocity profile. Data points show masers
and CO thermal emission from the wind. The SiO maser emission originates in the
extended atmosphere and is not part of the outflow.
}
\end{figure}

\subsection{Individual Sources}
Since \tV\ controls both the dynamics and the radiative transfer, in any given
source the spectral energy distribution (SED) can be fitted with only one free
parameter and the corresponding solution should provide a fit also for the
radial velocity profile. Figure 3 shows the SED fit for W Hya. A byproduct of
the solution that produced this SED is the shape of the dimensionless velocity
profile. There is no freedom to modify this profile, and all the data points
(from observations of CO thermal emission and various masers) indeed follow it
when scaled to dimensionless values. Figure 3 shows the results with actual
physical variables, setting the dimensional scales for $r$ and $v$ from the
Hipparcos distance of 115 pc and the terminal velocity of 8 \kms\ from CO
emission.

Significantly, neither $L$, \Mdot\ or dust-to-gas ratio were input parameters
of the model. Instead, these quantities are derived afterwards when the SED
fitting results are supplemented by the distance and velocity scales. The
bolometric flux is a direct product of the fitting process and together with
the distance to the source determines the overall luminosity. And once $L$, $v$
and \tV\ are known, \Mdot\ and the dust abundance are calculated from general
self-similarity relations.

\section{Water Thermal Emission}

Water is an abundant component of the outflow in oxygen-rich stars and the
dominant coolant at radial distances up to \about\ \E{15} cm. But because of
atmospheric absorption, the discovery of its thermal emission had to wait for
ISO. With the wind properties in W Hya determined from SED analysis, Zubko \&
Elitzur (2000) successfully fitted all 21 water lines observed by ISO using
\H2O abundance as the only free parameter. Harwitt \& Bergin (2002)
subsequently found the same line ratios in VY CMa. This highly luminous
supergiant has $L$ = 5\x\E5 \Lo\ and \Mdot\ $\simeq$ 2\x\E{-4} \Myr\ while in W
Hya the corresponding values are only 1.1\x\E4 \Lo\ and 2.3\x\E{-6} \Myr. In
spite of these large differences, scaling the model results for W Hya by an
overall factor of 10 produces adequate fit for the \H2O lines in VY Cma too.

Thanks to the multitude of observed transitions, in \H2O line analysis the
abundance is derived rather than entered as input. This is a significant
advantage over the common determination of \Mdot\ from CO observations, which
involves only one transition and must assume an abundance beforehand.

\begin{figure}
\includegraphics[width=0.99\hsize,clip]{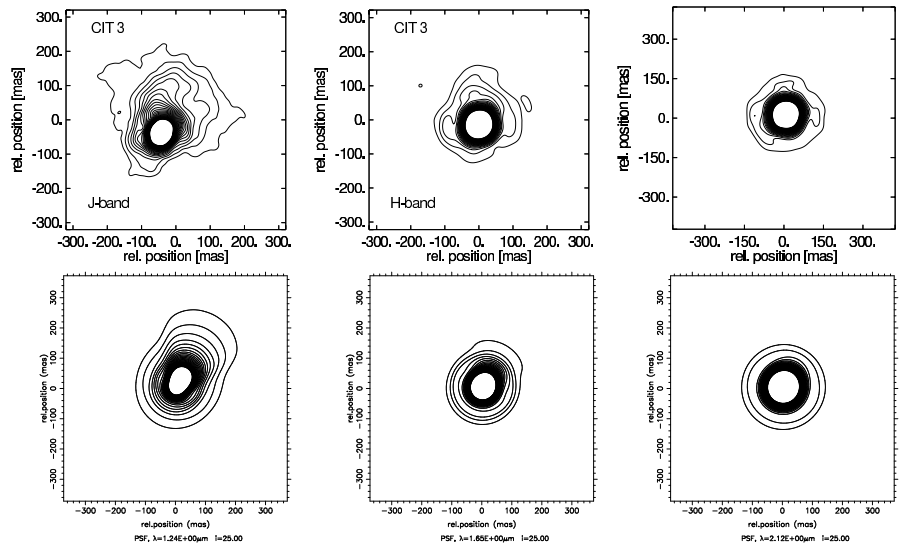}
\phantom{X.}
\includegraphics[width=0.28\hsize,clip]{moshe_fig4b.ps} \hskip 0.25in
\includegraphics[width=0.28\hsize,clip]{moshe_fig4c.ps} \hskip 0.22in
\includegraphics[width=0.28\hsize,clip]{moshe_fig4d.ps}
\caption{Top Row: Contour plots of the observed CIT3 images at 1.24 $\mu$m
(J-band), 1.65 $\mu$m (H-band), and 2.12 $\mu$m (K-band) from left to right
(Hofmann et al 2001). Bottom: Model results of Vinkovi\'c et al, 2002.}
\end{figure}

\section{Deviations from Spherical Symmetry}

Many AGB shells display spherical symmetry, but there are numerous exceptions.
A peculiar case is the source CIT3, one of the most extreme infrared AGB
objects. Observations by Hofmann et al (2001) show a 1.24\mic\ image elongated
along a symmetry axis (fig.\ 4, top left panel). The elongation disappears at
slightly longer wavelengths, and the image becomes almost perfectly circular at
2.12\mic. This peculiar behavior has been successfully explained by Vinkovi\'c
et al (2002) with the aid of a newly developed 2D radiative transfer code,
LELUYA (see www.leluya.org). In the model geometry, the wind spherical symmetry
is broken by two bi-polar cones and the observing viewpoint is inclined to the
axis. The density distributions in the conical regions and the bulk of the
outflow are deduced directly from the observed brightness of the J-band image.
The model results, shown in the lower panels of fig.\ 4, are hardly
distinguishable from the observations.

The wavelength-variation of the images reflects interplay between scattering
and emission. Scattered light dominates at $\lambda \la$ 1.5 \mic\ and the
images trace the large-scale contours of the density distribution. This
explains the elongated shape of the J-image. At longer wavelengths emission
takes over and the images are affected by the shape of the dust isotherms,
becoming circular due to the central heating of the dust.

\section{Conclusions}

The ``standard model'' seems to be working satisfactorily. Detailed fits are
successfully obtained with a single free parameter, \tV, as predicted. Dust
drift is a major ingredient of the dynamics and the reason for the peculiar
correlation discovered by Young (1995).

Two fundamental issues remain open and can be addressed only by more complete
models that incorporate the wind origin (such as those of the Berlin and Vienna
groups). The first is a more reliable determinations of the phase space lower
boundary shown in fig.\ 1. What is the minimal \Mdot? The second involves the
dust abundance, which sets the scale of terminal velocities in AGB outflows to
\about\ 10 \kms; were that abundance 100 times higher, typical velocities would
be \about\ 100 \kms\ instead. What is so special about $n_d\sigma_d/n$ \about\
\E{-22} cm$^{-2}$ and why is it the same in both C- and O-rich stars? It is
entirely possible that the answers to these two open issues are related.


\begin{chapthebibliography}{1}
\def\Ref{\bibitem{}}
\Ref Bowen, G.H. 1989, in {\em Evolution of peculiar red giant stars},
     ed. H.R. Johnson \& B. Zuckerman (Cambridge University Press: Cambridge),
     p. 269
\Ref Dorfi, E.A., H{\" o}fner, S., \& Feuchtinger, M.U.\ 2001, in
     {\em Stellar pulsation - nonlinear studies}, ed. M. Takeuti \& D.D.
     Sasselov (Dordrecht: Kluwer) p. 137
\Ref Elitzur, M., Brown, J.A. \& Johnson, H.R., 1989, ApJ, 341, L95
\Ref Elitzur, M., \& Ivezi\'c, Z. 2001, MNRAS, 327, 403
\Ref Gilman, R.C. 1972, ApJ, 178, 423
\Ref Goldreich, P., \& Scoville, N., 1976, ApJ, 205, 144 (GS)
\Ref Habing, H.J., Tignon, J. \& Tielens, A.G.G.M. 1994, A\&A 286, 523 (HTT)
\Ref Harwit, M. \& Bergin, E. A. 2002, ApJ Lett., 565, L105
\Ref Hofmann, K.-H., et al 2001, A\&A, 379, 529
\Ref Ivezi\'c \v Z., \& Elitzur M., 1995, ApJ, 445, 415
\Ref Ivezi\'c \v Z., \& Elitzur M., 2002, in preparation
\Ref Ivezi\' c, \v Z., Nenkova, M., \& Elitzur, M., 1999, User Manual for DUSTY,
     University of Kentucky Internal Report, accessible at
     http://www.pa.uky.edu/$\sim$moshe/dusty
\Ref Olofsson, H., Eriksson, K., Gustafsson, B., \& Carlstrom, U.
     1993, ApJS, 87, 267
\Ref Vinkovi\'c, D., Bloecker, T., Elitzur, M., Hofmann K.-H.,
     \& Weigelt, G., 2002, in preparations
\Ref Young, K. 1995, ApJ, 445, 872
\Ref Zubko, V.,  \& Elitzur, M. 2000, ApJ Lett., 544, L137


\end{chapthebibliography}
\end{document}